# Neuroaesthetics and the Science of Visual Experience

*Harish Vijayakumar, Design Solutions Engineer*
*Tamil Nadu, India. Email: thisisharish01@gmail.com*

## Abstract

Neuroaesthetics bridges neuroscience, psychology, and art to explore how our brains perceive and respond to visual beauty. This field seeks to demystify why certain designs or artworks "feel right," unraveling the neural mechanisms underlying our experience of visual pleasure. By examining the interplay between perception, emotion, and cognition, neuroaesthetics uncovers how beauty is constructed in the brain and how these insights can inform fields like graphic and interface design. This paper offers a comprehensive yet approachable overview of neuroaesthetics, aiming to make its key concepts accessible for anyone interested in the relationship between design, brain science, and human emotional engagement. The findings highlight that impactful design extends beyond mere aesthetics well-crafted visual experiences can connect, move, and support people in meaningful ways.

## 1. Introduction

Every day, we interact with visual designs whether it's a product label, a mobile app screen, or a painting on a wall. Sometimes, these visuals captivate us and feel "just right", even if we can't explain why. This intuitive sense of beauty, balance, or emotional pull is precisely what neuroaesthetics aims to understand.

Neuroaesthetics is a growing area of research that combines science and design. It explores how our brains respond to things we see and why certain visuals make us feel curious, moved, or calm. This field isn't just about art in galleries. It reaches into graphic design, UX/UI interfaces, branding, and even architecture any place where visual experience plays a role in how we think and feel. This paper introduces the basics of neuroaesthetics with a focus on practical applications in graphic design and digital interfaces. You don't need to be a neuroscientist or a professional designer to follow this the goal is to understand how the human brain experiences visual beauty, and how that knowledge can support better, more human-centered design.

## 2. What Is Neuroaesthetics?

Neuroaesthetics is a field at the intersection of neuroscience (how the brain works) and aesthetics (what we find beautiful or emotionally meaningful). Could there be specific patterns, colors, shapes, or layouts that the brain consistently responds to in a pleasing way? Neuroaesthetics aims to find out.

The field uses tools from brain science like EEG (electroencephalography) and fMRI (functional magnetic resonance imaging) to study what happens in the brain when we look at certain types of images, designs, or artworks.

While beauty is often considered subjective, research suggests that many people respond positively to things like balanced layouts, harmonious color palettes, and moderate complexity. These clues help designers create experiences that are more likely to feel enjoyable and intuitive to a wide audience.

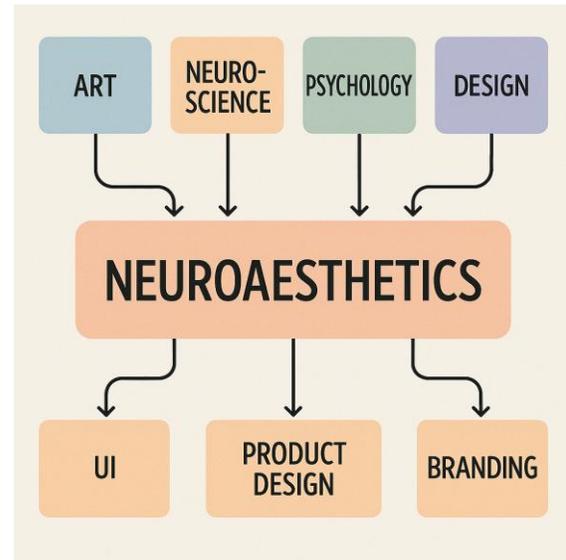

## 3. How the Brain Processes Visual Beauty

When we look at something, light reaches our eyes and travels into the brain's visual centers. But beauty and emotion go far beyond just "seeing."

Key Brain Areas Involved:

- Visual Cortex: Processes basic features like color, shape, and movement.

- Orbitofrontal Cortex: Linked to emotional reactions to beauty, especially feelings of pleasure or joy.

- Default Mode Network (DMN): Involved when we are deeply engaged with something emotional or personally meaningful like a powerful image, artwork, or memory.

- Amygdala and Limbic System: Play a role in emotional reactions like surprise, awe, or discomfort.

Together, these brain regions allow us to process not only what we're seeing but also how we feel about it and whether we want to keep looking or move on.

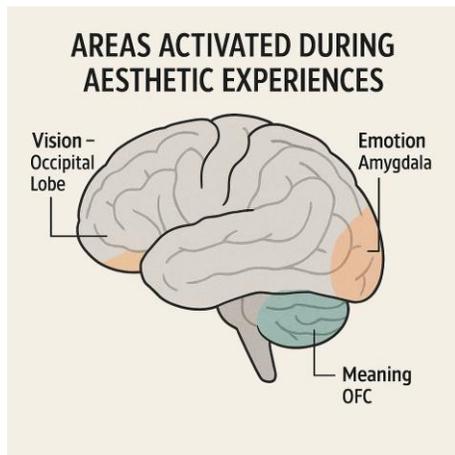

## 4. What Makes a Design Pleasing?

While individual taste does vary, certain design elements seem to consistently activate positive responses.

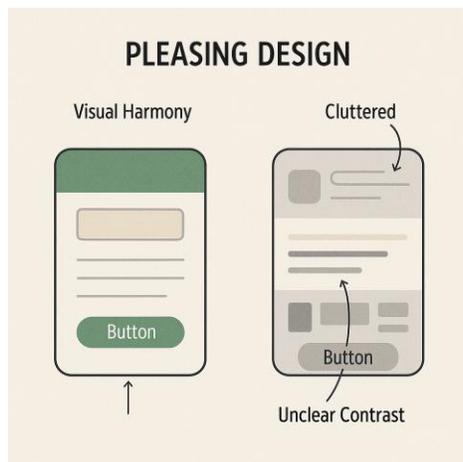

Common Features That Engage the Brain:

| Feature | Why It Works | Example in Design |
|---|---|---|
| Balance | Feels organized, calming | Symmetrical layouts in apps |
| Harmony | Reduces visual stress | Soft color palettes |
| Simplicity | Easier to process | Clean landing pages |
| Contrast | Guides attention, adds energy | Bold typography |
| Familiarity | Helps immediate recognition | Icon-based navigation |
| Novelty | Sparks interest, keeps eyes moving | Abstract illustrations |

The brain enjoys a challenge, but not too much. Designs that are slightly surprising or different, but still easy to understand, tend to be more engaging. Overly complicated or overly plain visuals often create discomfort or boredom.

## 5. How Neuroaesthetics Helps Designers

Graphic designers, UX/UI professionals, and digital artists can benefit from neuroaesthetic principles by using aesthetic choices that align with how users' brains work.

Key Applications in Design:

- User Interfaces: A balanced color scheme and well-structured layout help users process information faster. For example, people are more likely to trust websites that look professionally designed.
- Branding: Colors and shapes can influence brand perception. A circular logo feels more friendly than a sharp-edged one, for instance.
- Product Packaging: Designs that use symmetry, contrast, and focal points can improve shelf appeal and increase attention span in crowded spaces.
- Information Design: Well-designed typography and spacing improve readability and cognition, reducing mental fatigue when scrolling through apps or documents.

A Few Research-Based Tips:

- Use white space to prevent visual overload.
- Choose color palettes carefully as color directly influences emotion (blues are calming, reds are energizing).
- Keep layouts intuitive, especially in navigation, to reduce cognitive stress.
- Add microinteractions or animations that provide subtle feedback and increase satisfaction.

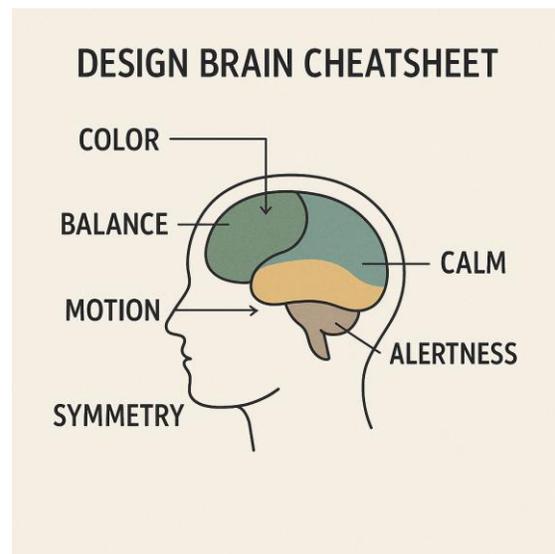

## 6. Cultural and Personal Differences

Beauty is both universal and personal. Some design preferences seem consistent across cultures, like appreciation for symmetry or natural color palettes.

However, many elements such as symbolism, color meaning, and aesthetic forms are shaped by upbringing, education, and culture.

Examples:

- White represents purity in many Western cultures but death in some Eastern traditions.
- Red can signal danger or celebration depending on context.
- Some cultures prefer minimalist design, others enjoy intricate patterns.

Designers should also consider how different users' brains perceive beauty and clarity, especially when creating accessible interfaces. People with neurodivergent conditions (such as ADHD or autism) may benefit from cleaner interfaces, fewer distractions, and controlled use of color and motion.

## 7. The Power of Visuals in Emotion and Memory

A well-designed visual experience does more than look good it creates emotional impact and lingers in memory. This is why design plays such a key role in storytelling, branding, and user engagement.

Aesthetically pleasing visuals:

- Activate memory centers, improving recall.
- Encourage positive emotional associations.
- Build trust and credibility.
- Increase time spent engaging with content.

When people say a brand feels "premium" or a website "just feels right," that's the result of intentional design working hand in hand with emotional and cognitive response.

## 8. Challenges and the Road Ahead

While neuroaesthetics is exciting, it's also still developing. Understanding the brain is complex, and no design can please everyone. Some ongoing challenges include:

- Individual differences: What works for one person might not for another.
- Data accuracy: Brain readings can be affected by many factors outside the visual experience.
- Ethics: Using brain science to influence behavior (especially in advertising or social apps) raises questions about manipulation.
-

- Accessibility: There's a long way to go in applying neuroaesthetics to make design truly inclusive for all users.

Despite these challenges, designers who stay curious and aware of how visual experience connects with emotion and cognition will be better equipped to design with empathy and impact.

## 9. Conclusion

Neuroaesthetics offers a fascinating look into the science behind why we find visual experiences enjoyable, meaningful, or powerful. It shows us that beauty is not just about looks it's a complex interaction between the brain, the eyes, and the emotions.

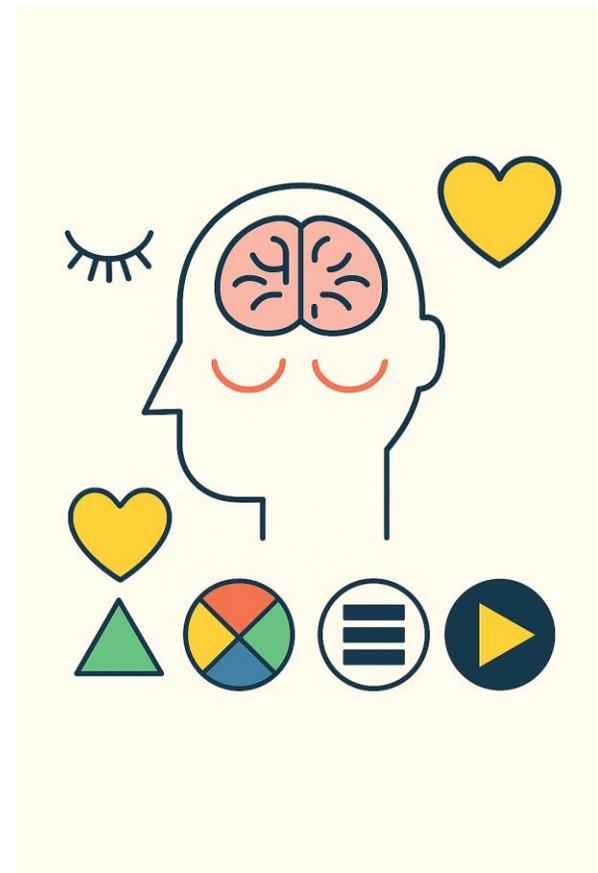

For designers, this means that every shape, color, layout, and animation has the potential to make a difference. By understanding how the brain responds to visuals, we can create designs that are not only visually appealing but also emotionally engaging, easier to use, and more human-centered.

In the end, great design is about more than just function or style. It's about understanding people, minds, and moments and neuroaesthetics helps build that bridge.

## References

[1] Chatterjee, A., & Vartanian, O. (2016). Neuroaesthetics: Brain mechanisms underlying aesthetic experience. *Nature Reviews Neuroscience, 17*(10), 685–696. https://doi.org/10.1038/nrn.2016.102

[2] Gerger, G., Leder, H., & Kremer, A. (2014). Context effects on emotional and aesthetic evaluations of artworks and non-art images. *Acta Psychologica, 151*, 174–

183. https://doi.org/10.1016/j.actpsy.2014.06.008

[3] Hsu, C.-T., Jacobs, A. M., & Conrad, M. (2015). The emotional power of poetry: Neural circuitry, psychophysiology and compositional principles. *Social Cognitive and Affective Neuroscience, 10*(1), 77–85. https://doi.org/10.1093/scan/nsu021

[4] Jacobsen, T., Schubotz, R. I., Höfel, L., & von Cramon, D. Y. (2006). Brain correlates of aesthetic judgment of beauty. *NeuroImage, 29*(1), 276–285. https://doi.org/10.1016/j.neuroimage.2005.07.010

[5] Leder, H., Belke, B., Oeberst, A., & Augustin, D. (2004). A model of aesthetic appreciation and aesthetic judgments. *British Journal of Psychology, 95*(4), 489–508. https://doi.org/10.1348/0007126042369811

[6] Nadal, M., & Skov, M. (2013). Introduction to the special issue: Toward an interdisciplinary neuroaesthetics. *Psychology of Aesthetics, Creativity, and the Arts, 7*(1), 1–12. https://doi.org/10.1037/a0031833

[7] Tschacher, W., Greenwood, S., Kirchberg, V., Wintzerith, S., van den Berg, K., & Tröndle, M. (2012). Physiological correlates of aesthetic perception of artworks in a museum. *Psychology of Aesthetics, Creativity, and the Arts, 6*(1), 96–103. https://doi.org/10.1037/a0023847

[8] Vessel, E. A., Starr, G. G., & Rubin, N. (2019). Art reaches within: Aesthetic experience, the self and the default mode network. *Frontiers in Neuroscience, 13*, 1176. https://doi.org/10.3389/fnins.2019.01176

[9] Zaidel, D. W. (2015). Neuroaesthetics is not just about art. *Frontiers in Human Neuroscience, 9*, 80. https://doi.org/10.3389/fnhum.2015.00080

[10] Zeki, S. (2001). Artistic creativity and the brain. *Science, 293*(5527), 51–52. https://doi.org/10.1126/science.1062331